\documentclass{article}
\usepackage{spconf,amsmath,graphicx}
\usepackage{booktabs}
\usepackage{amssymb}
\usepackage{bm}
\usepackage{multirow}
\usepackage{arydshln}
\usepackage{cite}
\usepackage{algorithm,algorithmicx,algpseudocode}
\usepackage{etoolbox,siunitx}
\usepackage{xcolor}
\usepackage{setspace}

\let\OLDthebibliography\thebibliography
\renewcommand\thebibliography[1]{
  \OLDthebibliography{#1}
  \setlength{\parskip}{.8pt}
  \setlength{\itemsep}{1pt plus 0.3ex}
}

\makeatletter
\def\adl@drawiv#1#2#3{%
        \hskip.5\tabcolsep
        \xleaders#3{#2.5\@tempdimb #1{1}#2.5\@tempdimb}%
                #2\z@ plus1fil minus1fil\relax
        \hskip.5\tabcolsep}
\newcommand{\cdashlinelr}[1]{%
  \noalign{\vskip\aboverulesep
           \global\let\@dashdrawstore\adl@draw
           \global\let\adl@draw\adl@drawiv}
  \cdashline{#1}
  \noalign{\global\let\adl@draw\@dashdrawstore
           \vskip\belowrulesep}}
\makeatother

\newcommand{\pz}{\phantom{0}}

\DeclareMathOperator*{\argmax}{argmax}

\title{Advancing Momentum Pseudo-Labeling with\\Conformer and Initialization Strategy}
\name{
    Yosuke Higuchi$^{1,2*}$\thanks{$^*$Work done during an internship at MERL},
    Niko Moritz$^{1}$,
    Jonathan Le Roux$^1$,
    Takaaki Hori$^1$
}

\address{
  $^1$Mitsubishi Electric Research Laboratories (MERL), USA\ \ $^2$Waseda University, Japan\\
  \small{\texttt{higuchi@pcl.cs.waseda.ac.jp, \{leroux, thori\}@merl.com}}
}

\begin{document}
\ninept
\maketitle
\setlength{\abovedisplayskip}{4pt}
\setlength{\belowdisplayskip}{4pt}
\setlength{\textfloatsep}{0.2cm} %
\begin{abstract}
\vspace{-1mm}
Pseudo-labeling (PL), a semi-supervised learning (SSL) method 
where a seed model performs self-training using pseudo-labels generated from untranscribed speech,
has been shown to enhance the performance of end-to-end automatic speech recognition (ASR).
Our prior work proposed momentum pseudo-labeling (MPL), which performs PL-based SSL via an interaction between online and offline models,
inspired by the mean teacher framework.
MPL achieves remarkable results on various semi-supervised settings,
showing robustness to variations in the amount of data and domain mismatch severity.
However, there is further room for improving the seed model used to initialize the MPL training,
as it is in general critical for a PL-based method to start training from high-quality pseudo-labels.
To this end,
we propose to enhance MPL by
(1) introducing the Conformer architecture to boost the overall recognition accuracy and
(2) exploiting iterative pseudo-labeling with a language model to improve the seed model before applying MPL.
The experimental results demonstrate that the proposed approaches effectively improve MPL performance,
outperforming other PL-based methods.
We also present in-depth investigations to make our improvements effective,
e.g., with regard to batch normalization typically used in Conformer and LM quality.

\end{abstract}
\vspace{-1mm}
\vspace{-2mm}
\section{Introduction}
\vspace{-2mm}
\label{sec:intro}
Recent progress in automatic speech recognition (ASR) has shifted towards the end-to-end (E2E) framework,
which aims to model direct speech-to-text conversion using a single deep neural network~\cite{graves2014towards, chorowski2015attention, chan2016listen}.
With well-established sequence-to-sequence modeling techniques~\cite{graves2006connectionist,graves2012sequence, sutskever2014sequence, bahdanau2014neural} 
and more sophisticated neural network architectures~\cite{dong2018speech,gulati2020conformer}, 
E2E ASR models have shown promising results on various benchmarks~\cite{chiu2018state, luscher2019rwth, karita2019comparative}.
However,
the performance often depends on the availability of a large quantity of labeled (transcribed) speech data,
which is not always feasible with high annotation costs.

To compensate for the limited amount of labeled data,
semi-supervised learning (SSL)~\cite{chapelle2009semi} methods that make use of a large amount of unlabeled data to improve the model performance can be applied.
While various efforts have been made to perform SSL in E2E ASR\cite{tjandra2017listening,hori2019cycle,ling2020deep,baevski2020wav2vec,zhang2020pushing,baskar2021eat},
a pseudo-labeling (PL)~\cite{lee2013pseudo} (or self-training~\cite{scudder1965probability})-based approach has been attracting increasing attention
due to its effectiveness and simplicity~\cite{kahn2020self,xu2020iterative,chen2020semi, park2020improved,khurana2021unsupervised,moritz2021semi,likhomanenko2021slimipl,higuchi2021momentum}.
In PL, 
a seed model is first trained on labeled data and used to generate pseudo-labels for unlabeled data.
Both the labeled and pseudo-labeled data are then used
to train a better-performing model.
In our previous work~\cite{higuchi2021momentum},
we proposed a PL-based method for E2E ASR, called momentum pseudo-labeling (MPL).
MPL trains a pair of online and offline models that interact and learn from each other,
inspired by the mean teacher framework~\cite{tarvainen2017mean}.
The online model is trained to predict pseudo-labels generated on the fly by the offline model, which maintains an exponential moving average of the online model parameters.
Through the interaction between the two models,
MPL effectively stabilizes the training with unlabeled data and
significantly improves over seed model performance.

One of the crucial factors for making PL-based approaches successful is
to avoid generating severely erroneous pseudo-labels,
which can lead to limiting the improvement of an E2E ASR model.
In a typical SSL setting in ASR,
the amount of labeled data is quite small and
the quality of pseudo-labels is not necessarily guaranteed.
To this end,
an external language model (LM) and beam-search decoding are often incorporated into the labeling process~\cite{kahn2020self,xu2020iterative}.
In~\cite{park2020improved,khurana2021unsupervised}, low-quality pseudo-labels are excluded via confidence-based filtering to promote model training with high-quality labels. And in~\cite{moritz2021semi},
an N-best list of pseudo-labels is leveraged to incorporate more appropriate supervision from alternative ASR hypotheses.

We believe that MPL still has room for further improvement by
making the models capable of generating pseudo-labels with higher quality.
Thus, in this work,
we propose to enhance MPL by
(1) introducing the Conformer architecture to boost the overall recognition accuracy, and
(2) using iterative pseudo-labeling~\cite{xu2020iterative} to transfer LM knowledge into a seed model before performing MPL.
The key contributions of this work are summarized as follows.
(a) We show that vanilla Conformer suffers from generalizing to unlabeled data,
especially when there is a domain mismatch against labeled data.
We mitigate this issue by substituting batch normalization with group normalization for the convolution module.
(b) We demonstrate that the improved MPL is robust to over-fitting to an LM training text set,
which has been reported as problematic for using an LM in PL~\cite{khurana2021unsupervised,likhomanenko2021slimipl}.
We also investigate the importance of LM quality in our framework.
(c) We show the proposed approaches effectively enhance MPL,
conducting experiments on a variety of SSL scenarios with varying amounts of unlabeled data or domain mismatch.

\vspace{-2mm}
\section{Momentum Pseudo-Labeling}
\vspace{-2mm}
\label{sec:mpl}
In this section,
we review the MPL method proposed in our prior work~\cite{higuchi2021momentum}.
MPL is described in two steps:
1) the supervised training of a seed E2E ASR model, and 
2) the MPL-based semi-supervised training of the model using unlabeled data.

\vspace{-3mm}
\subsection{Supervised training of a seed model}
\vspace{-1.2mm}
\label{ssec:mpl_sup}
E2E ASR is formulated as a sequence mapping problem between a $T$-length input sequence $X \!=\! (\bm{\mathrm{x}}_t \in \mathbb{R}^D| t\!=\!1,\dots,T)$ and 
an $L$-length output sequence $Y \!=\! ( y_l \in \mathcal{V} | l\!=\!1,\dots,L )$. 
Here, $\bm{\mathrm{x}}_t$ is a $D$-dimensional acoustic feature at frame $t$, 
$y_l$ an output token at position $l$, and $\mathcal{V}$ a vocabulary.
This work focuses on the connectionist temporal classification (CTC)-based E2E ASR model~\cite{graves2006connectionist, graves2014towards},
which is less prone to the looping and early stopping issues often caused by autoregressive decoder networks~\cite{chorowski2016towards, kahn2020self}.
CTC predicts a frame-level latent sequence $Z=(z_t \in \mathcal{V} \cup \{\epsilon\}| t=1,\dots,T)$, which is obtained by augmenting $Y$ with a
special blank token $\epsilon$.
Based on the conditional independence assumption between token predictions,
CTC models the conditional probability $P(Y|X)$ by marginalizing over latent sequences as
\begin{equation}
    \label{eq:p_ctc}
    P (Y | X) \approx \sum_{Z \in \mathcal{B}^{-1} (Y)} \prod_{t=1}^{T} P (z_t | X), 
\end{equation}
where $\mathcal{B}^{-1}(Y)$ returns all possible latent sequences compatible with $Y$.
Given labeled data $\mathcal{D}_{\mathsf{sup}} \!=\! \{ (X_n, Y_n) | n\!=\!1,\dots,N \}$, 
a seed model $P_\theta$ with parameters $\theta$ is optimized by minimizing the negative log-likelihood of Eq.~\eqref{eq:p_ctc}:
\begin{equation}
    \label{eq:l_sup}
    \mathcal{L}_{\mathsf{sup}} (\theta) = - \log P_{\theta}(Y_n | A(X_n)), 
\end{equation}
where $A(\cdot)$ indicates SpecAugment~\cite{park2019specaugment} for augmenting the input.

\vspace{-1.5mm}
\subsection{Semi-supervised training with MPL}
\vspace{-1.5mm}
\label{ssec:mpl_ssl}
The goal of semi-supervised training is to exploit unlabeled data $\mathcal{D}_{\mathsf{unsup}} \!=\! \{ X_m | m\!=\!N\!+\!1,\dots,N\!+\!M \}$ for enhancing the seed model trained on labeled data $\mathcal{D}_{\mathsf{sup}}$.
MPL performs the training using a pair of \textit{online} and \textit{offline} models that interact and learn from each other.
Let $P_{\xi}$ and $P_{\phi}$ be the online and offline models
with parameters $\xi$ and $\phi$,
which are initialized with the seed model parameters $\theta$.

\noindent\textbf{Online model training:}
Given an unlabeled sample $X\!\in\!\mathcal{D}_{\mathsf{unsup}}$,
the online model is trained on pseudo-labels $\hat{Y}$ generated on the fly by the offline model:
\begin{equation}
    \label{eq:y_hat}
    \hat{Y} = \argmax_{Y} P_{\phi} (Y | X), 
\end{equation}
where $\argmax$ is performed by the best path decoding of CTC~\cite{graves2006connectionist}. 
With the pseudo-labels generated from Eq.~\eqref{eq:y_hat}, 
the objective of the online model is defined in the same manner as Eq.~\eqref{eq:l_sup}:
\begin{equation}
    \label{eq:l_unsup}
    \mathcal{L}_{\mathsf{unsup}} (\xi) = - \log P_{\xi}(\hat{Y}_{N+m} | A(X_{N+m})),
\end{equation}
where $\mathcal{L}_{\mathsf{unsup}}$ is optimized via a gradient descent optimization.
Note that, during the semi-supervised training,
we also use labeled data $\mathcal{D}_{\mathsf{sup}}$ and the supervised loss  $\mathcal{L}_{\mathsf{sup}}(\xi)$,
which helps the online model stabilize and promote learning from unlabeled data with $\mathcal{L}_{\mathsf{unsup}}(\xi)$.

\noindent\textbf{Offline model training:}
After every update of the online model,
the offline model accumulates parameters of the online model via the momentum-based moving average:
\begin{equation}
    \label{eq:ema}
    \phi \leftarrow \alpha \phi + (1 - \alpha) \xi, 
\end{equation}
where $\alpha \!\in\! [0, 1]$ is a momentum coefficient.
This momentum update makes the offline model evolve more smoothly than the online model,
preventing the pseudo-labels from deviating too quickly from the labels initially generated by the seed model.
To handle the sensitive tuning of the momentum coefficient $\alpha$,
we follow our prior work and indirectly derive $\alpha$ from a weight $w \!=\! \alpha^K$,
where $K$ is the number of iterations (i.e., batches) in a training epoch.
The weight $w$ can be regarded as the proportion of the seed model retained after a training epoch, and
we fix it to 50\% (i.e., $w=0.5$),
as it has been shown consistently effective in various semi-supervised settings~\cite{higuchi2021momentum}.

\vspace{-2mm}
\section{Proposed Improvements for Momentum Pseudo-Labeling}
\vspace{-2mm}
We propose to enhance MPL by
(1) introducing the Conformer architecture~\cite{gulati2020conformer} to improve the overall accuracy,
and
(2) adopting iterative pseudo-labeling (IPL)~\cite{xu2020iterative} to transfer LM knowledge into the seed model.
We expect these approaches to promote the MPL training
by enabling the models to generate higher-quality pseudo-labels.

\vspace{-1.5mm}
\subsection{Conformer for semi-supervised training}
\vspace{-1.5mm}
Conformer is a variant of Transformer augmented with convolution to increase the capability for capturing local feature patterns~\cite{gulati2020conformer}.
In addition to the multi-head self-attention layer in the Transformer encoder,
Conformer introduces a module based on depthwise separable convolution~\cite{chollet2017xception}.
Unlike Transformer,
Conformer employs relative positional encoding and macaron-like feed-forward layers.

While Conformer-based models have achieved outstanding ASR performance compared with standard Transformers~\cite{guo2021recent},
we empirically observe that Conformer suffers from poor generalization from labeled data to unlabeled data.
A similar issue has been reported in other ASR tasks~\cite{li2021better,liu2021end,kim2021generalizing}.
Simply adopting Conformer for MPL makes the training become unstable and diverge easily,
especially when a domain mismatch exists between labeled and unlabeled data.

We assume that such a problem comes from unreliable statistics computed and used by batch normalization (BN)~\cite{ioffe2015batch} in the convolution module.
As we suppose the amount of labeled data relatively small (i.e., 100h),
the estimated mean and variance of the whole dataset are likely to
become less accurate in BN~\cite{ioffe2017batch}.
A simple solution is to increase the mini-batch size.
However, we observe that a large mini-batch size degrades the seed model performance,
which can lead to degrading the quality of pseudo-labels during the MPL training.
Hence,
we consider replacing BN with group normalization (GN)~\cite{wu2018group} in the convolution module,
as it has been investigated in~\cite{kriman2020quartznet,li2021better}.
GN divides feature maps into groups and
normalizes the features within each group,
which makes the training less dependent on the mini-batch size.
This is found critical for stabilizing the Conformer-based MPL training,
as examined in Sec.~\ref{exp:conformer}.

\begin{algorithm}[t]
    \renewcommand{\ttdefault}{cmtt}
    \caption{\bf Momentum pseudo-labeling using iterative pseudo-labeling for transferring LM knowledge into seed model}
    \label{algo:mpl+ipl}
    \begin{algorithmic}[1]
    \scriptsize
    \renewcommand{\algorithmicrequire}{\textbf{Input:}}
    \renewcommand{\algorithmicensure}{\textbf{Output:}}
        \Statex \textbf{Input:}
        \Statex \ \ \ $\mathcal{D}_{\mathsf{sup}}, \mathcal{D}_{\mathsf{unsup}}$ \Comment{labeled and unlabeled data}
        \Statex \ \ \ $\mathcal{A}$  \Comment{an ASR model architecture}
        \Statex \ \ \ $\alpha$ \Comment{a momentum coefficient}
        \Statex \vspace{-7pt}
        \State {\scriptsize\ttfamily\textcolor{blue}{\# 1.Seed model training}}
        \State Train a seed model $P_{\theta}$ with architecture $\mathcal{A}$ on $\mathcal{D}_{\mathsf{sup}}$ using Eq.~\eqref{eq:l_sup}
        \Statex \vspace{-7pt}
        \State {\scriptsize\ttfamily\textcolor{blue}{\# 2.Iterative pseudo-labeling}}
        \For {$i=1,...,I_{\mathsf{ipl}}$}
            \State Generate pseudo-labels $\hat{\mathcal{D}}_{\mathsf{unsup}} = \{ (X_m, \hat{Y}_m) | X_m \in \mathcal{D}_{\mathsf{unsup}}\}$,
            \Statex \qquad\quad using $P_\theta$ and LM with beam-search decoding
            \For {$e=1,...,E_{\mathsf{ipl}}$}
                \ForAll {$(X, Y) \in \mathcal{D}_{\mathsf{sup}} \cup \hat{\mathcal{D}}_{\mathsf{unsup}}$}
                    \State Compute loss $\mathcal{L}$ for $P_{\theta}(Y|X)$ with Eq.~\eqref{eq:l_sup}
                    \State Update $\theta$ using ${\nabla}_{\theta}\mathcal{L}$
                \EndFor
            \EndFor
        \EndFor
        \Statex \vspace{-7pt}
        \State {\scriptsize\ttfamily\textcolor{blue}{\# 3.Momentum pseudo-labeling}}
        \State Initialize an online model $P_{\xi}$ and an offline model $P_{\phi}$ with $P_{\theta}$
        \For {$e=1,...,E_{\mathsf{mpl}}$}
            \ForAll {$S \in \mathcal{D}_{\mathsf{sup}} \cup \mathcal{D}_{\mathsf{unsup}}$}
                \State Obtain $X \sim S$
                \State Obtain $Y = \begin{cases}
                    Y \sim S & (S \in \mathcal{D}_{\mathsf{sup}})\\
                    \hat{Y} \sim P_{\phi} (Y | X) & (S \in \mathcal{D}_{\mathsf{unsup}})
                \end{cases}$
                \State Compute loss $\mathcal{L}$ for $P_{\xi}(Y|X)$ with Eq.~\eqref{eq:l_sup} or~\eqref{eq:l_unsup}
                \State Update $\xi$ using ${\nabla}_{\xi}\mathcal{L}$
                \State Update $\phi \leftarrow \alpha \phi + (1 - \alpha) \xi$
            \EndFor
        \EndFor \\
        \Return $P_\xi$ \Comment{online model is returned for final evaluation}
    \end{algorithmic}
\end{algorithm}

\vspace{-2mm}
\subsection{Iterative pseudo-labeling for enhancing seed model}
\vspace{-1.5mm}
To provide the MPL training with a better model for initializing the online and offline models,
we consider enhancing the seed model using (IPL).
IPL continuously trains a model with periodic regeneration of pseudo-labels,
where an external LM and beam-search decoding are used
to generate the labels~\cite{xu2020iterative}.
While beam-search decoding with an LM plays an important role for generating pseudo-labels with high quality~\cite{kahn2020self,wallington2021on},
it is computationally intensive for MPL due to the on-the-fly label generation.
Hence,
we exploit IPL to implicitly transfer LM knowledge to the seed model before applying MPL,
providing the MPL training with a better initialization for generating higher-quality pseudo-labels.
Moreover,
by not using the LM-based pseudo-labels during the MPL traning,
we can prevent the model from over-fitting to the LM training text data,
which often degrades the generalization capability of the ASR model~\cite{khurana2021unsupervised,likhomanenko2021slimipl}.

Algorithm~\ref{algo:mpl+ipl} shows the proposed MPL training with IPL initialization.
In the beginning, a seed model is trained using a labeled set as in Sec.~\ref{ssec:mpl_sup} (line 1--2).
Then,
the seed model is further trained via IPL with LM and beam-search decoding (line 3--12).
Here, we denote $I_{\mathsf{ipl}}$ as the number of iterations (pseudo-label updates), and 
$E_{\mathsf{ipl}}$ as the number of epochs trained in each iteration.
We refer to standard pseudo-labeling (PL)~\cite{kahn2020self} when $I_{\mathsf{ipl}}\!=\!1$ and IPL~\cite{xu2020iterative} when $I_{\mathsf{ipl}}\!>\!1$.
Finally,
the enhanced seed model is used to initialize the models for MPL (line 13--23).
The MPL training lasts $E_{\mathsf{mpl}}$ epochs.

In our prior work,
we have discussed a little about applying PL as the initialization strategy for MPL~\cite{higuchi2021momentum} and
demonstrated its effectiveness.
This work extends this early idea by focusing on the better-performing IPL.
In Sec.~\ref{ssec:analysis_lm}, we also investigate the influence of the quality of LM used for PL on improving MPL.

\vspace{-2mm}
\section{Experiments}
\vspace{-2mm}

\subsection{Experimental setting}
\vspace{-1.5mm}
\noindent\textbf{Data:}
We conducted experiments using the LibriSpeech (LS)~\cite{panayotov2015librispeech} and TEDLIUM3 (TED3)~\cite{hernandez2018ted} datasets.
LS is a corpus of read English speech,
containing 960 hours of training data (split into ``train-clean-100'', ``train-clean-360'', and ``train-other-500'').
TED3 is a corpus of English Ted Talks consisting of 450 hours of training data (``train-ted3'').
We used the standard development and test sets for each dataset.
Kaldi~\cite{povey2011kaldi} was used to extract
80 mel-scale filterbank coefficients with three-dimensional pitch features.
For text tokenization,
we used a 1k subword vocabulary,
which was constructed from the ``train-clean-100'' transcriptions using SentencePiece~\cite{kudo2018subword}.

\noindent\textbf{Semi-supervised settings:} 
After training a seed model on the labeled ``train-clean-100'' (LS-100) set,
we considered three semi-supervised settings with different unlabeled sets:
LS-100/LS-360, 
an in-domain setting with ``train-clean-360'' (LS-360); 
LS-100/LS-860, 
an in-domain setting with ``train-\{clean-360,other-500\}'' (LS-860); and 
LS-100/TED3, 
an out-of-domain setting with ``train-ted3''. 

\noindent\textbf{ASR model:}
We used the Conformer architecture~\cite{gulati2020conformer} implemented in ESPnet~\cite{watanabe2018espnet},
which consists of two convolutional neural network layers followed by a stack of 12 self-attention layers.
The number of heads $H$, dimension of a self-attention layer $d_{\mathsf{model}}$, 
dimension of a feed-forward network $d_{\mathsf{ff}}$, and kernel size $K$ were set to 
4, 256, 2048, and 31, respectively.
We set the number of groups to 8 for group normalization when used in the convolution module.

\noindent\textbf{Training configuration:}
We basically followed our prior work~\cite{higuchi2021momentum}.
The seed model was trained for 150 epochs using the Adam optimizer~\cite{kingma2015adam},
and Noam learning rate scheduling~\cite{vaswani2017attention}.
The semi-supervised training was done up to 200 epochs,
where the gradient-based optimization was done by using
the Adam optimizer.
IPL was performed by iterating PL for the maximum of 8 times ($I_{\mathsf{ipl}}\!\le\!8$),
where the model was trained for 25 epochs ($E_{\mathsf{ipl}}\!=\!25$) in each iteration.
Note that, after each iteration, we averaged model parameters over the last 5 checkpoints
to stabilize the pseudo-label generation.
We set $w\!=\!0.5$ for MPL training, following our prior work~\cite{higuchi2021momentum}.

\noindent\textbf{Decoding configuration:}
For evaluation, a final model was obtained by averaging model parameters over 10 checkpoints with the best validation performance.
We trained an LM consisting of 4 long short-term memory (LSTM) layers with 2048 units,
using the LS-100 transcriptions combined with the external text data provided by LibriSpeech~\cite{panayotov2015librispeech}.
For decoding with the LM,
we adopted a frame-synchronous CTC prefix beam search algorithm~\cite{hannun2014first,moritz2019streaming},
where we used a beam-size of 20,
a score-based pruning threshold of 14.0,
an LM weight of 1.0, and
an insertion bonus factor of 2.0.
For decoding without an LM,
we performed the best path decoding of CTC~\cite{graves2006connectionist}.

\begin{table}[t]
    \centering
    \caption{Validation WER [\%] for seed models trained on labeled LS-100. For the Conformer-based models, we explored different normalization methods for the convolution module.}
    \label{tb:conformer}
    \renewcommand{\arraystretch}{0.85}
    \scalebox{0.85}{
    \begin{tabular}{llccc}
        \toprule
        & & \multicolumn{2}{c}{\textbf{LibriSpeech}} & \textbf{TED3} \\
        \cmidrule(l{0.3em}r{0.3em}){3-4}\cmidrule(l{0.3em}r{0.3em}){5-0}
        \textbf{Model} & \textbf{Norm.\ type} & dev-clean & dev-other & Dev \\
        \midrule
        Transformer & -- & 12.2 & 30.0 & 31.2 \\
        \midrule
        \multirow{4}{*}[0pt]{Conformer} & Batch & \pz8.6 & 23.1 & 27.3 \\
        & Instance & \pz8.9 & 23.5 & 27.1 \\
        & Group & \textbf{\pz8.4} & \textbf{22.5} & \textbf{26.4} \\
        & Layer & \textbf{\pz8.4} & 22.9 & 26.9 \\
        \bottomrule
    \end{tabular}}\vspace{-0.1cm}
\end{table}

\begin{figure}[t]
    \centering
    \includegraphics[width=0.89\columnwidth]{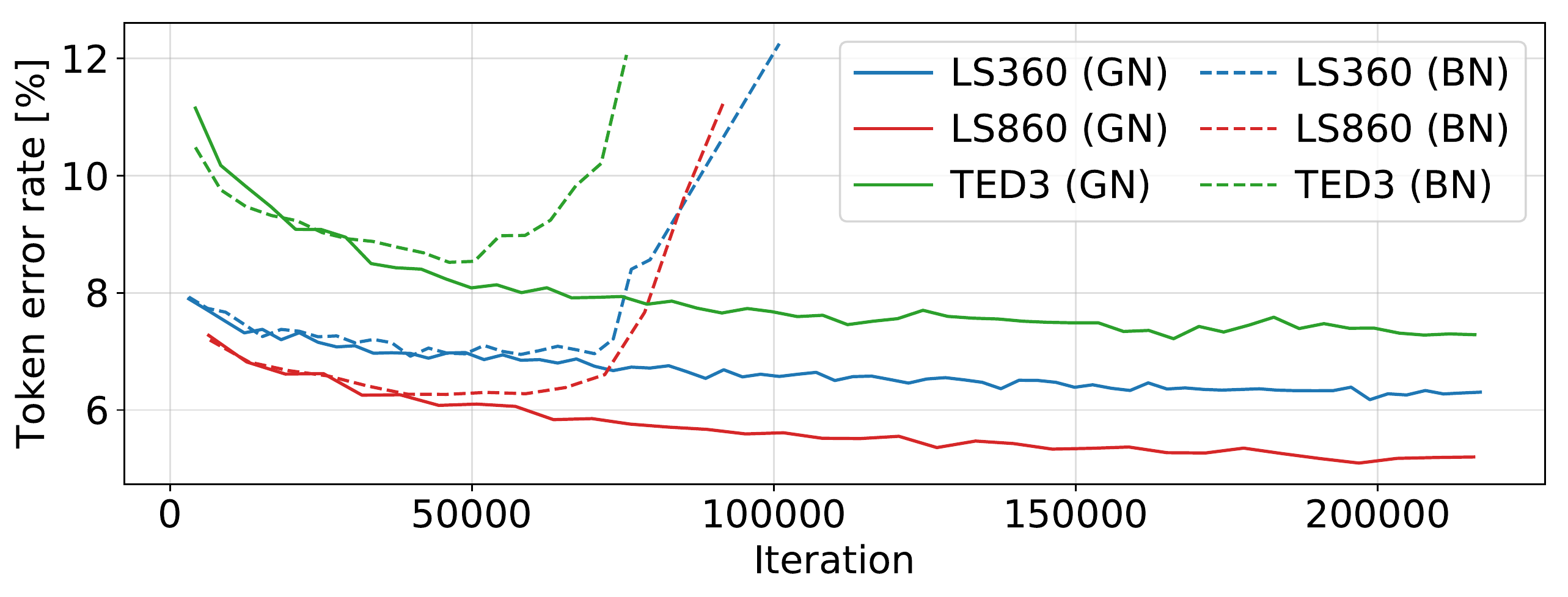}
    \vspace{-5mm}
    \caption{Validation token error rate [\%] of MPL training using Conformer with batch (dotted line) or group (solid line) normalization.}
    \label{fig:conformer_mpl}
\end{figure}

\begin{table*}[t]
    \centering
    \caption{Word error rate (WER) [{\footnotesize $\%$}] and WER recovery rate (WRR) [{\footnotesize $\%$}] on in-domain LibriSpeech (LS) settings. 
    The results are divided into two sections: whether the LM with beam-search decoding was applied in the \underline{final evaluation} or not. $\dagger$ indicates trained for 100 epochs.}
    \label{tb:ls}
    \scalebox{0.85}{
    \renewcommand{\arraystretch}{0.85}
    \begin{tabular}{lllcccccccccccc}
        \toprule
        & & & \multicolumn{6}{c}{Decoding without LM} & \multicolumn{6}{c}{Decoding with LM} \\
        \cmidrule(l{0.3em}r{0.3em}){4-9} \cmidrule(l{0.3em}r{0.8em}){10-15}
        & & & \multicolumn{2}{c}{\textbf{Dev WER}} & \multicolumn{2}{c}{\textbf{Test WER}} & \multicolumn{2}{c}{\textbf{Test WRR}} & \multicolumn{2}{c}{\textbf{Dev WER}} & \multicolumn{2}{c}{\textbf{Test WER}} & \multicolumn{2}{c}{\textbf{Test WRR}} \\
        \cmidrule(l{0.3em}r{0.3em}){4-5}
        \cmidrule(l{0.3em}r{0.3em}){6-7} \cmidrule(l{0.3em}r{0.3em}){8-9} \cmidrule(l{0.3em}r{0.3em}){10-11} \cmidrule(l{0.3em}r{0.3em}){12-13} \cmidrule(l{0.3em}r{0.3em}){14-15} 
        & \textbf{Method} & \textbf{Init.} & clean & other & clean & other & clean & other & clean & other & clean & other & clean & other \\
        \midrule
        LS-100 & \texttt{L0}\hspace{1mm} seed (Cfm) & -- & \pz8.4 & 22.5 & \pz8.6 & 23.3 & \pz\pz0.0 & \pz\pz0.0 & \pz5.2 & 15.2 & \pz5.5 & 16.0 & \pz\pz0.0 & \pz\pz0.0 \\
        \midrule
        \multirow{7}{*}[-5pt]{\shortstack[l]{LS-100\\\ \ / LS-360}} & \texttt{A0}\hspace{1mm} MPL (Trf) & seed (Trf) & \pz8.7 & 21.4 & \pz9.0 & 21.7 & \pz-- & \pz-- & \pz4.8 & 13.0 & \pz5.1 & 13.1 & \pz-- & \pz-- \\
        &\texttt{A1}\hspace{1mm} MPL & \texttt{L0} & \pz6.1 & 16.0 & \pz6.6 & 15.8 & \pz52.3 & \pz76.4 & \pz4.5 & 11.2 & \pz4.7 & \textbf{11.1} & \pz34.7 & \pz71.7 \\
        & \texttt{A2}\hspace{1mm} PL & \texttt{L0} & \pz5.7 & 15.9 & \pz6.1 & 15.8 & \pz64.6 & \pz76.0 & \pz4.3 & 11.4 & \pz4.5 & 11.8 & \pz40.6 & \pz62.2 \\
        & \texttt{A3}\hspace{1mm} IPL & \texttt{L0} & \textbf{\pz5.4} & 15.1 & \pz5.7 & 15.3 & \pz73.3 & \pz81.5 & \pz4.2 & 11.5 & \pz4.5 & 11.7 & \pz42.2 & \pz62.5 \\
        & \texttt{A4}\hspace{1mm} MPL$^{\dagger}$ & \texttt{A2}@ep100 & \pz5.7 & 15.5 & \pz6.1 & 15.6 & \pz64.8 & \pz77.8 & \pz4.2 & 11.1 & \pz4.5 & 11.3 & \pz44.0 & \pz69.3 \\
        & \texttt{A5}\hspace{1mm} MPL$^{\dagger}$ & \texttt{A3}@ep100 & \pz5.5 & \textbf{15.0} & \textbf{\pz5.6} & \textbf{15.1} & \textbf{\pz75.1} & \textbf{\pz83.3} & \textbf{\pz4.1} & \textbf{10.8} & \textbf{\pz4.3} & \textbf{11.1} & \textbf{\pz51.4} & \textbf{\pz72.7} \\
        \cdashlinelr{2-15}
        & \texttt{A6}\hspace{1mm} topline & \texttt{L0} & \pz4.1 & 13.6 & \pz4.7 & 13.4 & 100.0 & 100.0 & \pz2.9 & \pz9.4 & \pz3.2 & \pz9.2 & 100.0 & 100.0 \\
        \midrule
        \multirow{7}{*}[-5pt]{\shortstack[l]{LS-100\\\ \ / LS-860}} & \texttt{B0}\hspace{1mm} MPL & seed (Trf) & \pz8.1 & 16.5 & \pz8.3 & 16.8 & \pz-- & \pz-- & \pz4.6 & \pz9.7 & \pz4.8 & 10.1 & \pz-- & \pz-- \\
        & \texttt{B1}\hspace{1mm} MPL & \texttt{L0} & \pz5.7 & 12.2 & \pz6.2 & 12.2 & \pz48.1 & \pz76.4 & \pz4.1 & \pz8.5 & \pz4.4 & \pz8.7 & \pz36.5 & \pz74.6 \\
        & \texttt{B2}\hspace{1mm} PL & \texttt{L0} & \pz5.4 & 13.9 & \pz5.7 & 14.2 & \pz57.8 & \pz62.3 & \pz4.0 & 10.5 & \pz4.2 & 10.7 & \pz43.0 & \pz53.7 \\
        & \texttt{B3}\hspace{1mm} IPL & \texttt{L0} & \textbf{\pz4.7} & 11.5 & \textbf{\pz5.0} & 11.7 & \textbf{\pz71.0} & \pz79.3 & \pz4.1 & \pz9.7 & \pz4.4 & 10.2 & \pz36.2 & 	\pz58.9\\
        & \texttt{B4}\hspace{1mm} MPL$^{\dagger}$ & \texttt{B2}@ep100 & \pz5.1 & 12.1 & \pz5.3 & 12.4 & \pz64.0 & \pz75.1 & \pz3.7 & \pz8.4 & \pz3.9 & \pz8.8 & \pz51.0 & \pz73.3 \\
        & \texttt{B5}\hspace{1mm} MPL$^{\dagger}$ & \texttt{B3}@ep100 & \textbf{\pz4.7} & \textbf{11.0} & \textbf{\pz5.0} & \textbf{11.1} & \pz70.0 & \textbf{\pz83.9} & \textbf{\pz3.6} & \textbf{\pz7.8} & \textbf{\pz3.8} & \textbf{\pz8.2} & \textbf{\pz54.0} & \textbf{\pz79.6} \\
        \cdashlinelr{2-15}
        & \texttt{B6}\hspace{1mm} topline & \texttt{L0} & \pz3.3 & \pz9.0 & \pz3.5 & \pz8.7 & 100.0 & 100.0 & \pz2.4 & \pz6.1 & \pz2.5 & \pz6.2 & 100.0 & 100.0 \\
        \bottomrule
    \end{tabular}}\vspace{-0.5cm}
\end{table*}

\begin{table*}[t]
    \centering
    \caption{WER [{\footnotesize $\%$}] and WRR [{\footnotesize $\%$}] on out-domain TEDLIUM3 (TED3) setting.}
    \centering
    \label{tb:ted3}
    \scalebox{0.85}{
    \renewcommand{\arraystretch}{0.85}
    \begin{tabular}{lllcccccc}
    \toprule
    & & & \multicolumn{3}{c}{Decoding without LM} & \multicolumn{3}{c}{Decoding with LM} \\ 
    \cmidrule(lr{0.5em}){4-6} \cmidrule(l{-0.1em}r{0.5em}){7-9}
    \textbf{Setting} & \textbf{Method} & \textbf{Init.} & \textbf{Dev WER} & \textbf{Test WER} & \textbf{Test WRR} & \textbf{Dev WER} & \textbf{Test WER} & \textbf{Test WRR} \\
    \midrule
    LS-100 & \texttt{L0}\hspace{1mm} seed (Cfm) & -- & 26.4 & 26.5 & \pz\pz0.0 & 21.3 & 21.1 & \pz\pz0.0 \\
    \midrule
    \multirow{5}{*}[-5pt]{\shortstack[l]{LS-100\\\ \ / TED3}} & \texttt{C0}\hspace{1mm} MPL (Trf) & seed (Trf) & 18.4 & 17.0 & \pz-- & 14.9 & 13.3 & \pz-- \\
    & \texttt{C1}\hspace{1mm} MPL & \texttt{L0} & 15.1 & 13.9 & \pz81.0 & 12.7 & \textbf{11.6} & \textbf{\pz77.3} \\
    & \texttt{C2}\hspace{1mm} IPL & \texttt{L0} & 16.8 & 16.8 & \pz62.2 & 16.6 & 16.9 & \pz34.2 \\
    & \texttt{C3}\hspace{1mm} MPL$^\dagger$ & \texttt{C2}@ep100 & \textbf{14.6} & \textbf{13.8} & \textbf{\pz81.1} & \textbf{12.4} & 12.0 & \pz73.8 \\
    \cdashlinelr{2-9}
    & \texttt{C4}\hspace{1mm} topline & \texttt{L0} & 10.4 & 10.9 & 100.0 & \pz8.6 & \pz8.8 & 100.0 \\
    \bottomrule
    \end{tabular}}\vspace{-0.5cm}
\end{table*}

\vspace{-2.8mm}
\subsection{Effectiveness of adopting Conformer for MPL}
\vspace{-1.2mm}
\label{exp:conformer}
In Table~\ref{tb:conformer},
we compare the word error rate (WER) of seed models trained with the Transformer (Trf) or the Conformer (Cfm) architecture.
For Cfm-based models, we investigated different normalization methods for the convolution module,
including \{batch~\cite{ioffe2015batch}, instance~\cite{ulyanov2016instance}, group~\cite{wu2018group}, layer~\cite{ba2016layer}\} normalization (\{BN, IN, GN, LN\}).
Note that IN and LN are the same as GN with group sizes 1 and 256 ($=\!d_{\mathsf{model}}$), respectively.
Comparing the two architectures,
the Cfm-based models significantly improved over the Trf-based model.
Within the Cfm-based models,
GN resulted in the best performance on both LS and TED3, and
the 100-hour training data seemed to be too small to take advantage of BN.
As normalizing across feature maps (i.e., IN, GN, and LN) achieved better performance than BN on the out-of-domain TED3 set,
it indicates that BN led to lower generalization capability with unreliable statistics.
Note that in~\cite{kriman2020quartznet},
BN achieved better performance than the other normalization methods
when another depthwise separable convolution-based ASR model is trained on the full 960-hour set of LS.

Figure~\ref{fig:conformer_mpl} shows learning curves from MPL training using Cfm with BN or GN.
In all semi-supervised settings, BN caused the training to become unstable.
Especially in the out-of-domain setting with TED3,
the model diverged more quickly than in the other settings.
In contrast, GN successfully stabilized the MPL training with Cfm.

\vspace{-2.8mm}
\subsection{Results on in-domain setting}
\vspace{-1.2mm}
Table \ref{tb:ls} lists results on the in-domain LS settings
in terms of the WER and WER recovery rate (WRR)~\cite{ma2008unsupervised}.
The topline results were obtained via fully supervised training on each setting.
Looking at the MPL results (\texttt{A1},\texttt{B1}),
MPL led to a substantial improvement over the seed model (\texttt{L0}),
effectively learning from unlabeled data using Cfm with GN.
These Cfm results significantly outperformed those of prior Trf-based MPL~\cite{higuchi2021momentum} (\texttt{A0},\texttt{B0} vs.\ \texttt{A1},\texttt{B1}).
With pseudo-labels generated using the LM,
PL~\cite{kahn2020self} and IPL~\cite{xu2020iterative} achieved lower WERs on the ``clean'' sets
than those obtained from MPL, and
IPL resulted in better performance than MPL on the ``other'' sets as well (\texttt{*2},\texttt{*3} vs.\ \texttt{*1}).
However, when decoded with the LM,
the performance gain was larger for MPL with a slight decrease in WRRs, and
MPL achieved much lower WERs on the ``other'' sets.
PL and IPL, in contrast, had smaller improvement with degraded WRRs,
which indicates PL and IPL are fitted to LM knowledge and have less variations in the hypotheses during the beam-search decoding.
\texttt{*4} and \texttt{*5} show results for the proposed MPL training using the seed model enhanced by PL and IPL, respectively.
Note that we performed PL or IPL for 100 epochs and MPL for another 100 epochs to
match the total training epochs of the other methods.
The initialization strategy provided MPL with distinct improvements,
pushing the limit of the other methods (\texttt{*4},\texttt{*5} vs.\ \texttt{*1},\texttt{*2},\texttt{*3}).
With the IPL-based initialization, MPL achieved the best overall performance on both of the settings with different amounts of unlabeled data (\texttt{A5},\texttt{B5}).
Moreover,
when decoded with the LM,
the improved MPL retained higher WRRs than IPL (\texttt{*3} vs.\ \texttt{*5}),
maintaining the advantage of MPL and
making the model less dependent on the LM knowledege.

\vspace{-3mm}
\subsection{Results on out-of-domain setting}
\vspace{-1.2mm}
Table \ref{tb:ted3} shows MPL results on the TED3 setting.
Cfm with GN significantly improved MPL over the seed model and Trf-based MPL (\texttt{C1} vs.\ \texttt{L0},\texttt{C0}),
successfully stabilizing training on the out-of-domain data.
IPL led to a decent improvement over the seed model, but
the gain was more substantial for MPL (\texttt{C1} vs.\ \texttt{C2}).
As there is a domain mismatch between the LM training text and
the actual transcriptions of TED3,
IPL was less effective at learning from the out-of-domain unlabeled data.
Moreover, IPL had little gain from decoding with the LM,
indicating the model was prone to over-fit to the LM knowledge.
By using IPL to enhance the seed model,
MPL further reduced WERs (\texttt{C1} vs.\ \texttt{C3}).
However,
the improvement was much smaller than those observed in the in-domain settings, and
the standard MPL performed sufficiently well by decoding with the LM.

\vspace{-1.5mm}
\subsection{Does better language model lead to better MPL results?}
\vspace{-1.2mm}
\label{ssec:analysis_lm}
\begin{table}[t]
    \centering
    \vspace{-2mm}
    \caption{WER for MPL initialized by PL with different LM quality.}
    \centering
    \label{tb:analysis_lm}
    \renewcommand{\arraystretch}{0.85}
    \scalebox{0.85}{
    \begin{tabular}{lcc@{$\ \rightarrow\ $}cc@{$\ \rightarrow\ $}c}
    \toprule
    & & \multicolumn{2}{c}{Small LM} & \multicolumn{2}{c}{Large LM} \\
    \cmidrule(lr{0.5em}){3-4} \cmidrule(lr{0.5em}){5-6}
    \textbf{Setting} & \textbf{Test data}& \textbf{PL} & \textbf{MPL} & \textbf{PL} & \textbf{MPL} \\
    \midrule
    \multirow{2}{*}[0pt]{LS-100 / LS-360} & test-clean
    & \multicolumn{1}{c@{\phantom{$\ \rightarrow\ $}}}{\pz6.3} & \pz6.2
    & \multicolumn{1}{c@{\phantom{$\ \rightarrow\ $}}}{\pz6.2} & \pz6.1 \\
    & test-other
    & \multicolumn{1}{c@{\phantom{$\ \rightarrow\ $}}}{16.8} & 15.9
    & \multicolumn{1}{c@{\phantom{$\ \rightarrow\ $}}}{16.4} & 15.6 \\
    \midrule
    \multirow{2}{*}[0pt]{LS-100 / LS-860} & test-clean
    & \multicolumn{1}{c@{\phantom{$\ \rightarrow\ $}}}{\pz6.2} & \pz5.7
    & \multicolumn{1}{c@{\phantom{$\ \rightarrow\ $}}}{\pz5.7} & \pz5.3 \\
    & test-other
    & \multicolumn{1}{c@{\phantom{$\ \rightarrow\ $}}}{15.3} & 12.9
    & \multicolumn{1}{c@{\phantom{$\ \rightarrow\ $}}}{14.5} & 12.4 \\
    \bottomrule
    \end{tabular}}
\end{table}
In Table \ref{tb:analysis_lm}, we study the importance of LM quality used in PL to improve MPL performance.
We focus on in-domain settings (\texttt{A4},\texttt{B4} in Table~\ref{tb:ls}),
where the initialization strategy was especially effective.
We compare a small LM (1-layer LSTM) and large LM (4-layer LSTM),
validation perplexities of which were 20.9 and 14.3, respectively.
PL was evaluated at epoch 100,
which is then used to initialize MPL.
As a result, the large LM led to better PL performance and, accordingly,
improved MPL with better pseudo-label generation.

\vspace{-2mm}
\section{Conclusions}
\vspace{-2mm}
We proposed several improvements to momentum pseudo-labeling (MPL) for semi-supervised ASR.
Experimental results on various semi-supervised settings demonstrated
the effectiveness of the enhanced MPL,
showing clear improvements over our prior results and other PL-based methods.
Moreover, we investigated and shared the key components to make the proposed approaches effective,
including normalization method for Conformer and
the quality of LM for generating pseudo-labels.
Future work should consider evaluating MPL on lower-resource scenarios (e.g., 10h of labeled data~\cite{kahn2020libri}).

\newpage

\fontsize{8.3pt}{0pt}\selectfont
\bibliographystyle{IEEEtran}
\bibliography{refs}

\end{document}